\documentstyle[12pt]{article}

\def\be{\begin{equation}}
\def\ee{\end{equation}}
\def\bea{\begin{eqnarray}}
\def\eea{\end{eqnarray}}
\def\ov{\overline}
\def\dl{\delta}
\def\p{\psi_{-}}

\begin{document}
\vspace*{2cm}
\begin{center}
{\large\bf ASPECTS OF T-DUALITY IN OPEN STRINGS}
\vskip 1.5 cm
{\bf Javier Borlaf}\footnote{\tt javier@delta.ft.uam.es}
\vskip 0.05cm
Departamento de F\'{\i}sica Te\'orica, Universidad Aut\'onoma,
28049 Madrid, Spain
\vskip 0.9cm
{\bf Yolanda Lozano}\footnote{\tt yolanda@puhep1.princeton.edu}
\vskip 0.05cm
Joseph Henry Laboratories, Princeton University,
Princeton NJ 08544, USA
\end{center}

\date{ }
\setcounter{page}{0} \pagestyle{empty}
\thispagestyle{empty}
\vskip 2 cm
\begin{abstract}
We study T-duality for open strings in various $D$-manifolds
in the approach of
canonical transformations. We show that this approach is
particularly useful to study the mapping of the boundary
conditions since it provides an explicit relation between
initial and dual variables.
We consider non-abelian duality transformations and show
that under some restrictions the dual is a curved 
$(d-{\rm dim}G-1)$ D-brane, where $d$ is the dimension of the
space-time and $G$ the non-abelian symmetry group.
The generalization to $N=1$ supersymmetric sigma models with
abelian and non-abelian isometries is also considered.
\end{abstract}
\vfill
\begin{flushleft}
FTUAM-96-26, PUPT-1633\\
hep-th/9607051\\
July 1996
\end{flushleft}
\newpage\pagestyle{plain}

\def\theequation{\thesection . \arabic{equation}}

\section{Introduction}
\setcounter{equation}{0}

Recently there has been renewed interest in the study of 
open string theories
with the last developments in string dualities
(see for instance \cite{P} and references therein).
In \cite{Pol} Polchinski showed that open strings with 
certain exotic boundary conditions (D-branes) were the 
carriers
of the RR charges required by string duality \cite{var}
being the quantum of the charge precisely that predicted
by duality \cite{HSVW}.
This identification allowed for many new tests of
string duality. For instance the
heterotic string soliton needed by
the S-duality conjecture has been identified as a D-string 
in the type I superstring \cite{PW}. 
D-brane technology has also provided a statistical interpretation
of Black Hole entropy.
For very pedagogical reviews on D-branes and complete set
of references see \cite{PCJ}.

D-branes first arised as particular features under 
T-duality in
theories of open strings \cite{DLP,HG}. The
well-known duality mapping of toroidal compactifications:
\be
\label{p1}
\partial_+ x\rightarrow -\frac{1}{R^2}\partial_+
{\tilde x}, \qquad
\partial_- x\rightarrow \frac{1}{R^2}\partial_-{\tilde x}
\ee
maps Neumann boundary conditions:
$\partial_n x=0$, to Dirichlet boundary conditions:
$\partial_t {\tilde x}=0$, where $\partial_n$ and $\partial_t$ are
the normal and tangent derivatives to the boundary.
The ends of the strings are then confined to the 
${\tilde x}$ plane, which
is itself dynamical.
These particular objects with mixed Neumann and Dirichlet
boundary conditions are the D-branes \cite{DLP}.
For type I superstrings 
crosscap boundary conditions for the unoriented topologies are
mapped to orientifold conditions \cite{DLP,orien} and 
the dual D-brane
is hidden in the orientifold plane.
These dual theories may seem rather exotic, but they are just
a more suitable description at small distances
of the same original open string theory.

The open string-D-brane dualities of toroidal
compactifications have been
extended recently to more general backgrounds. Namely, to
backgrounds with abelian \cite{ABB,DO,BR,GHT} and
non-abelian isometries
\cite{FKS}. In these references the gauging procedure
to T-duality (for a review see \cite{GPR})
was followed\footnote{In \cite{DO} there is also a brief
study with canonical transformations.}.
Backgrounds without isometries have also been studied
in \cite{KS} within the Poisson-Lie T-duality.

In this article we study T-duality for open strings in various
backgrounds within the canonical transformation approach
\cite{AAL,GRVMV}. This approach is
particularly useful in obtaining
information about the boundary conditions,
since it provides an explicit mapping between initial and 
dual variables. In particular,
it generalizes (\ref{p1}) to the specific type of 
sigma model and duality symmetry 
under consideration. Once the duality mapping is identified
we show how it modifies the boundary conditions.
For backgrounds with abelian isometries
we will reproduce in a very simple manner
some of the results already presented in the references above.
In the non-abelian case we will see that the canonical
transformation gives results differing from the
ones in \cite{FKS}. We will give some arguments in
favor of the canonical transformation description 
of non-abelian duality for D-manifolds,
basically due to the existence of boundaries.

In section 2 we study
abelian T-duality of bosonic open and closed strings.
We reproduce in a very simple way the results in \cite{ABB,DO}. 
If the initial $d$ dimensional theory satisfies Neumann conditions 
then the generalization of (\ref{p1}) obtained with the
canonical transformation yields Dirichlet boundary
conditions for one of the coordinates (for backgrounds with just one
abelian isometry) and Neumann for the rest, i.e. the dual theory
is a $(d-2)$ D-brane. 
For backgrounds of unoriented strings it is also
easy to see that crosscap boundary conditions are mapped to
orientifold conditions \cite{ABB}.  

In section 3 we generalize these results to the case of $N=1$ open
superstring sigma models. With the canonical transformation
description we are able to consider
the most general case of
classical boundary conditions (minimizing both the bulk and the
boundary) and explore their transformation under abelian T-duality.
We show that in order to get a dual super D-brane some restrictions
over the original backgrounds must be impossed. For standard
Neumann R-NS boundary conditions we recover the results in
\cite{ABB}.

In section 4 we consider backgrounds with non-abelian
isometries. We argue that the
canonical transformation approach is more adequate
for open string world-sheets since the existence of boundaries 
makes unclear 
the validity of 
the more conventional gauging procedure. 
We focus in the particular class
of backgrounds with non-abelian isometries for which the canonical
transformation description is known, i.e. the ones where the
isometry group acts without isotropy.
We show that only in those cases in which the equivalence
between the initial and dual theories can be established at the
quantum level\footnote{We will clearly specify what we mean
by this.}, the dual of Neumann boundary conditions for the
coordinates transforming under the non-abelian symmetry group are
generalized Dirichlet boundary conditions, generalized in the
sense that it is the momentum defined in a given curved background
which is zero at the ends of the string. 
The inert coordinates
under the non-abelian isometry still satisfy generalized Neumann 
conditions in the dual.
Therefore the dual is a curved $(d-{\rm dim}G-1)$ D-brane, 
where $G$ is the non-abelian symmetry group,
static, as we will show.
We will see that for unoriented
strings crosscap boundary conditions are mapped to generalized
(in the same sense as above) orientifold conditions.
The D-branes are then hidden in the orientifold surfaces.
These results mean that flat D-branes and orientifolds 
are just particular
results under T-duality. 
We also present the $N=1$ supersymmetrization of some of the
models considered.

\section{Abelian Duality}
\setcounter{equation}{0}

Let us consider open and closed strings propagating in a $d$ dimensional
background
of metric, antisymmetric tensor and abelian gauge field\footnote{We
only consider throughout the paper abelian background
gauge fields. For a non-abelian treatment see \cite{DO}.}.
In the neutral
case the action can be written:
\be
\label{1uno}
S=\int_\Sigma d\sigma_+ d\sigma_-(g_{ij}+b_{ij})
\partial_+ x^i \partial_- x^j+\int_{\partial\Sigma}
V_i \partial_t x^i 
\ee
where $V_i$ denotes the abelian background gauge field and
$\partial_t$ is the tangent derivative to the 
boundary\footnote{We consider $\sigma=$ constant boundaries
throughout the paper but in certain, specified, cases.}.
The boundary term
can be absorbed in the action by just considering:
\be
\label{1dos}
S=\int_\Sigma d\sigma_+ d\sigma_- (g_{ij}+B_{ij})
\partial_+ x^i \partial_- x^j
\ee
with $B_{ij}=b_{ij}+F_{ij}=b_{ij}+\partial_i V_j-\partial_j V_i$.
The torsion term is absent for the unoriented topologies.
Let us assume that there exists a Killing vector $k^i$ such that
${\cal L}_k g_{ij}=0$ and ${\cal L}_k B_{ij}=0$ (this means we can
have: ${\cal L}_k b_{ij}=\partial_i v_j-\partial_j v_i$ and
${\cal L}_k V_i=-v_i+\partial_i\varphi$, for some $v_i$, 
$\varphi$). 
Going to the system of
adapted coordinates to the isometry $\{\theta,x^\alpha\}$,
$\alpha=1,\dots,d-1$, such that
$k=\partial_\theta$ the action reads:
\be
\label{1tres}
S=\frac12\int_{\Sigma}d^2\sigma (g_{00} \partial_+ \theta 
\partial_-\theta +
(g_{0\alpha}+B_{0\alpha})\partial_+\theta\partial_- x^\alpha+
(g_{0\alpha}-B_{0\alpha})\partial_+ x^\alpha\partial_-\theta+
(g_{\alpha\beta}+B_{\alpha\beta})\partial_+ x^\alpha\partial_-
x^\beta ).
\ee
Let us briefly discuss the gauging procedure to 
construct the abelian T-dual \cite{RV}.
This method has been applied to open strings in \cite{ABB,DO}.
The global isometry of the model can be made local by replacing 
ordinary derivatives
of the adapted coordinate by covariant derivatives, introducing a gauge
field $A_\pm$ with gauge variation $\delta A_\pm=-\partial_\pm \epsilon$
($\epsilon$ being the gauge parameter). This gauge field is imposed to
be flat by means of a Lagrange multiplier term. In
non-trivial world-sheets (with boundaries or non-trivial genera) 
one has to be
especially careful in imposing this condition. In \cite{AABL} the higher
genus case was studied in detail for closed strings. It was shown that
in order to obtain
$A$ pure gauge in non-trivial world-sheets the Lagrange
multiplier term must be introduced as
\be
\label{1tres1}
\int_{\Sigma} d{\tilde \theta}\wedge A
\ee
where ${\tilde \theta}$ must be multivalued to accomplish gauge invariance
under large gauge transformations. This multivaluedness can be accounted for
with a harmonic contribution to $d{\tilde \theta}$.
The integration on the exact component fixes the constraint that
$A$ is closed and, further, the integration on the harmonic component
imposes that the
harmonic contribution to $A$ is cero.
For open strings we have to consider as well the harmonic contribution
from the boundary, whose integration imposes the constraint 
\be
\label{1tres11}
\oint_{\partial\Sigma}A=0.
\ee
In this way the pure gauge condition
$A_\pm=\partial_\pm \theta_0$ is obtained. 
Doing the transformation $\theta\rightarrow\theta+\theta_0$
the original theory is recovered in gauge invariant variables
(recall: $\delta\theta=\epsilon, \delta A_\pm=-\partial_\pm \epsilon
\Rightarrow \delta\theta_0=-\epsilon$). On the other hand integrating
out $A_\pm$ and fixing the gauge $\theta=0$ the dual theory is obtained.
The dual backgrounds are given by Buscher's
formulas \cite{BU}:
\bea
\label{1cuatro1}
&&{\tilde g}_{00}=\frac{1}{g_{00}}, \quad
{\tilde G}_{0\alpha}=-\frac{B_{0\alpha}}{g_{00}},\quad
{\tilde B}_{0\alpha}=-\frac{g_{0\alpha}}{g_{00}},\,\,\,\nonumber\\
&&{\tilde G}_{\alpha\beta}=g_{\alpha\beta}-\frac{g_{0\alpha}
g_{0\beta}-B_{0\alpha}B_{0\beta}}{g_{00}},\quad
{\tilde B}_{\alpha\beta}=B_{\alpha\beta}-\frac{g_{0\alpha}B_{0\beta}-
g_{0\beta}B_{0\alpha}}{g_{00}}
\eea
where now $B$ involves the background gauge field 
and we have denoted the dual backgrounds also with capital
letters to account for this dependence.
${\tilde \theta}$ plays now the role of 0-coordinate. 
The regularization of the determinant coming from the integration
over the gauge fields produces the shift of the dilaton required
by conformal invariance \cite{BU,AO}:
${\tilde \Phi}=\Phi-\log{\rm det}g_{00}$.
In \cite{ABB,DO}
due account was taken  
of the boundary conditions with the result that the dual of (\ref{1uno})
with Neumann conditions in the boundary is a Dirichlet $(d-2)$-brane.

For closed strings it is also possible to construct the dual by the
method of the canonical transformation \cite{AAL,GRVMV}. 
The mapping:
\bea
\label{1cuatro}
&&p_\theta=- {\tilde \theta}^\prime \nonumber\\
&&p_{{\tilde \theta}}=-{\theta}^\prime
\eea
from $\{\theta, p_\theta\}$ to $\{{\tilde \theta}, p_{{\tilde \theta}}\}$,
where $p_\theta, p_{{\tilde \theta}}$ are the conjugate momenta
to $\theta, {\tilde \theta}$, yields the dual theory defined
by Buscher's formulas.
This relation provides as well the generalization of the non-local
duality transformation $d\rightarrow \frac{1}{R^2} *d$ of toroidal
compactifications to arbitrary backgrounds with abelian isometries.
We just need to substitute:
\bea
\label{1seis}
&&p_\theta=g_{00}{\dot \theta}+g_{0\alpha}{\dot x}^\alpha
-b_{0\alpha}x^{\prime\alpha} \nonumber\\
&&p_{\tilde \theta}=\frac{1}{g_{00}}(\dot{\tilde{\theta}}-
b_{0\alpha}{\dot{x}}^\alpha+g_{0\alpha}
x^{\prime\alpha} )
\eea
in (\ref{1cuatro}) to obtain:
\bea
\label{141}
&&\partial_+\theta=-{\tilde g}_{00}\partial_+{\tilde \theta}-
{\tilde k}^-_\alpha\partial_+{\tilde x}^\alpha=
-{\tilde k}^-_i\partial_+{\tilde x}^i
 \nonumber\\
&&\partial_-\theta={\tilde g}_{00}\partial_-{\tilde \theta}+
{\tilde k}^+_\alpha\partial_-{\tilde x}^\alpha={\tilde k}^+_i
\partial_-{\tilde x}^i
\eea
where we have defined $k^\pm_i=g_{0i}\pm b_{0i}$.

The generating functional  
is given by:
\be
\label{1cinco}
{\cal F}=\frac12\oint d\sigma ({\tilde \theta}{\theta}^\prime-
{\tilde \theta}^\prime \theta).
\ee
${\cal F}$ being linear in $\theta$ and ${\tilde \theta}$ implies that
the classical canonical transformation (\ref{1cuatro}) is also valid
quantum mechanically and we can write the relation:
\be
\label{1cinco1}
|{\tilde \theta}\rangle=\int {\cal D}\theta(\sigma)
e^{i{\cal F}[{\tilde \theta},
\theta(\sigma)]} |\theta (\sigma)\rangle
\ee
up to some normalization factor, between the corresponding Hilbert
spaces \cite{GC}. However renormalization effects still need to be
considered and in fact there are some results showing that they
give corrections to Buscher's backgrounds \cite{correc}. Also
to really establish the quantum equivalence between the initial
and dual theories we need to reproduce the dilaton shift within
the canonical transformations approach. Consider a constant
toroidal background of radius $R$. The measure in configuration
space is given by ${\cal D}\theta {\rm det}R$. We can regularize
this determinant 
as $R^{B_0}$, where
$B_0$ is the dimension of the space of 0-forms in the two 
dimensional world-sheet (regularized in a lattice, for instance).
With this prescription\footnote{This way of regularizing the
determinants has been shown to reproduce the correct modular
anomaly under S-duality in abelian gauge theories
\cite{Wi,BY}.}
one realizes that the usual
measure in phase space:
${\cal D}\theta{\cal D}p_\theta$ gives upon integration in
$p$: ${\cal D}\theta R^{B_1}$, where $B_1$ is the dimension of 
the space of 1-forms in the world-sheet and emerges because
the momenta are 1-forms. Therefore it differs from our
definition of measure in configuration space. In order to
reproduce the partition function in configuration space we
need to include explicit factors on $R$ in the definition of
the measure in phase space. One can check that considering these
factors the correct shift of the dilaton is obtained after
performing the canonical transformation. These arguments 
however are only rigorous for constant backgrounds. We believe
that a similar reasoning could be applied to the general case.

The relation (\ref{1cinco1}) between the states of the initial
and dual theories implies that
the duality transformation is valid for arbitrary 
Riemann surfaces, because $|\theta(\sigma)\rangle$ can be the
result of integrating the original theory on an arbitrary
Riemann surface with boundary.
Then the previous results apply with no modification 
to open and closed
strings backgrounds \cite{DO} with $b$ replaced
by $B$ in order to absorb the background gauge field.
In this case we also need to care about the boundary conditions.
The canonical transformation approach is particularly adequate to deal
with boundary conditions since it provides an
explicit relation between the target space
coordinates of the original and dual theories. 
{}From (\ref{141})
we get:
\be
\label{1nueve}
{\dot {\tilde \theta}}=-(g_{00}\theta^\prime +g_{0\alpha}
{\theta^\prime}^\alpha -B_{0\alpha} {\dot x}^\alpha)
\ee
and
\be
\label{1diez}
g_{0\alpha}\theta^\prime+g_{\alpha\beta}x^{\prime\beta}+
B_{0\alpha}{\dot \theta}-B_{\alpha\beta}{\dot x}^\beta= 
{\tilde G}_{0\alpha}{\tilde \theta}^\prime+{\tilde G}_{\alpha\beta}
x^{\prime\beta}+{\tilde B}_{0\alpha}{\dot {\tilde \theta}}-
{\tilde B}_{\alpha\beta}{\dot x}^\beta .
\ee
Then, Neumann boundary conditions for the original theory 
\cite{CLNY}:
\be
\label{1siete}
g_{ij}x^{\prime j}-B_{ij}{\dot x}^j=0
\ee
imply:
\bea
\label{1once}
&&{\dot {\tilde \theta}}=0 \nonumber\\
&&{\tilde G}_{0\alpha}{\tilde \theta}^\prime
+{\tilde G}_{\alpha\beta}
x^{\prime\beta}+{\tilde B}_{0\alpha}{\dot {\tilde \theta}}-
{\tilde B}_{\alpha\beta}{\dot x}^\beta=0.
\eea
These mixed boundary conditions represent a flat Dirichlet 
$(d-2)$-brane in
the dual background. Also from ({\ref{141}) we can deduce the collective
motion of the brane. Decomposing 
$B_{0\alpha}=b_{0\alpha}-\partial_\alpha V_0$ we realize that the
usual Buscher's backgrounds for closed strings (with the torsion
$b$) are gotten provided we redefine 
${\tilde \theta}\equiv {\tilde \theta}+V_0(x^\alpha)$.
Therefore $V_0(x^\alpha)$ gives the transverse position of the
brane in the dual theory\footnote{A particular case is when
$V_0$ is taken pure gauge locally breaking $U(N)$ to $U(1)^N$,
i.e. when a Wilson line 
$V_0={\rm diag}\{\theta_1,\dots,\theta_N\}$ is included.
In this case we get a maximum of $N$ D-branes in the dual
theory with fixed positions at $\theta_i$, $i=1,\dots,N$
\cite{PCJ}.}.
If we dualize $n$ commuting 
isometries it is straightforward to check that a  Dirichlet 
$(d-n-1)$-brane is obtained in the dual. 
It is perhaps worth mentioning that there are some particular backgrounds
(those whose conserved currents associated to the isometry are chiral
\cite{RV}) which are at the same time backgrounds of open strings and
D-branes depending on the boundary conditions, which are in turn
related by a
T-duality transformation.

In \cite{ABB} this mapping from Neumann to Dirichlet boundary conditions
was studied within the gauging procedure. 
As an intermediate stage
it was necessary to generalize Neumann boundary conditions for
the gauged action and the choice they made was to impose (\ref{1siete})
with covariant derivatives replacing ordinary derivatives of the adapted
coordinate. The justification for this is that the integration over
the Lagrange multiplier yields $A_\pm=\partial_\pm
\theta_0$, pure gauge, and after substituting in the covariant derivatives
and making the isometric transformation $\theta\rightarrow \theta+\theta_0$
the original Neumann boundary conditions in gauge invariant variables
are recovered. On the other hand,
substituting $A_\pm$ from their equations of motion and fixing the gauge
chosing $\theta=0$ Dirichlet boundary conditions are obtained for 
${\tilde \theta}$ and
Neumann for the rest of the coordinates. In the canonical transformation
approach we have made use of the explicit relations between 
original and dual
variables and further justified the choice made in \cite{ABB}.

Let us finish this section by analyzing the unoriented topologies.
Invariance under world-sheet parity implies that the
antisymmetric tensor and the abelian gauge field are projected
out of the spectrum. We can still have non-abelian gauge fields
in $SO(N)$ and $USp(N)$ but they must be treated differently
(see for instance \cite{DO}).
Unoriented topologies
can be obtained from oriented ones by identifications
of points on the boundary \cite{GSW}. For instance the projective
plane is obtained from the disk\footnote{We have to make first a
Wick rotation to imaginary time.}
identifying opposite points. The 
topology thus obtained is a crosscap.  
Under abelian T-duality we should get the mapping from crosscap
to orientifold conditions \cite{ABB}.
Crosscap boundary conditions for the coordinate 
adapted to the isometry:
\bea
\label{1doce}
&&{\dot \theta}(\sigma+\pi)=-{\dot \theta}(\sigma)\nonumber\\
&&\theta^\prime(\sigma+\pi)=\theta^\prime(\sigma),
\eea
where we are parametrizing the boundary of the disk by
$(0,2\pi)$ and identifying opposite points:
$\theta(\sigma+\pi)=\theta(\sigma)$,
translate to:
\bea
\label{1trece}
&&p_\theta(\sigma+\pi)=-p_\theta(\sigma)\nonumber\\
&&\theta^\prime (\sigma+\pi)=\theta^\prime (\sigma)
\eea
in phase space. Then
(\ref{1cuatro}) implies:
\bea
\label{1catorce}
&&{\tilde \theta}^\prime (\sigma+\pi)=
-{\tilde \theta}^\prime (\sigma) \nonumber\\
&&p_{{\tilde \theta}}(\sigma+\pi)=p_{{\tilde \theta}}(\sigma),
\eea
which are orientifold conditions in phase space since
$p_{{\tilde \theta}}(\sigma+\pi)=p_{{\tilde \theta}}(\sigma)$
implies ${\dot {\tilde \theta}}(\sigma+\pi)={\dot 
{\tilde \theta}}(\sigma)$\footnote{Integration on the first 
equation implies ${\tilde \theta}(\sigma+\pi)=
-{\tilde \theta}(\sigma)$ so that the world-sheet 
parity reversal is accompanied by a $Z_2$ transformation in
space-time. These kinds of constructions are the orientifolds
\cite{DLP,orien}.}.
The orientifold plane is at ${\tilde \theta}=0$ and it's
non-dynamical, because the abelian
gauge field is zero in the unoriented case.
These conditions can also be easily
worked out in configuration space from (\ref{141}).
The rest of the coordinates still
satisfy crosscap boundary conditions. 
This implies that we cannot have
${\tilde g}_{0i}$ components in the dual metric, which it can be
seen to be the case since $B_{0i}$ must be zero in the initial
theory.

Abelian T-duality has also been studied at the level of the
effective world volume actions in \cite{B,ABB,BR,GHT}.

\section{Superstrings}
\setcounter{equation}{0}

In this section we analyze abelian T-duality for $N=1$ open 
superstring sigma models
in the approach of canonical transformations.
This approach has been applied to closed superstrings
in \cite{H}.
The action in $(1,1)$ superspace reads:
\be
\label{(1.1)}
S = \int_{\Sigma} d\sigma_+d\sigma_- d^2\theta 
(g_{ij}+B_{ij})D_{+}X^iD_{-}X^j
\ee
where the superfields $X^i(\theta_{\pm},\sigma^{\pm}) = x^i + 
\theta_{+}\psi_{-}^i + \theta_{-}\psi_{+}^i + 
\theta_{+}\theta_{-}F^i$, the superderivatives $D_{+} = 
\partial_{\theta_{-}}-i\theta_{-}\partial_{\sigma_{+}}$ ,  $D_{-} = 
-\partial_{\theta_{+}} +i \theta_{+}\partial_{\sigma_{-}}$ and 
$B_{ij} = b_{ij} + F_{ij}$, as in the previous section.
Once the superspace integration $\int d^2\theta = 
D_{+}D_{-}\vert_{\theta_{\pm}=0}$ and the on-shell 
substitution of the auxiliar 
field $F^i$ are made, we get the usual target-space 
action's covariant form 
in components in addition to a boundary contribution 
which includes the supersymmetric 
circulation of the abelian electromagnetic field:
\bea
\label{(1.2)}
&&S=\int_{\Sigma} d\sigma_+ d\sigma_- \{(g_{ij} + 
b_{ij})\partial_{+}x^i\partial_{-}x^j - 
i g_{ij}\psi_{+}^i\nabla_{-}^{(+)}\psi_{+}^j - 
i g_{ij}\psi_{-}^i\nabla_{+}^{(-)}\psi_{-}^j+ \nonumber\\ 
&&{1\over2}R_{ijkl}(\Gamma^{(-)})\psi_{+}^i
\psi_{+}^j\psi_{-}^k\psi_{-}^l\} +
\int_{\partial\Sigma}\{V_i \partial_t x^i+\frac{i}{4}
B_{ij}(\psi_+^i\psi_+^j+\psi_-^i\psi_-^j)\}
\eea
where
\bea
\label{(1.3)}
&&\nabla^{(\pm)}_{\mp}\psi^i = \partial_\mp\psi^i + 
\partial_{\mp} 
x^j\Gamma_{jk}^{i(\pm)}\psi^k  \nonumber\\
&&\Gamma_{jk}^{i(\pm)} =\Gamma_{jk}^i \pm 
H_{jk}^i\nonumber\\
&&H_{ijk} ={1\over2}(b_{ij,k} + b_{jk,i} - 
b_{ik,j})
\eea
and $\Gamma^i_{jk}$ is the Levi-Civita connection.

For our purposes it is preferable to absorb the boundary term
in the bulk and work with an action:
\bea
\label{(1.3.1)}
&&S=\int_{\Sigma}d\sigma_+d\sigma_-\{(g_{ij}+
B_{ij})\partial_+
x^i\partial_-x^j-i\psi_{+}^i(g_{ij}+ B_{ij})
\partial_- \psi_+ ^j-i\psi_-^i(g_{ij}-B_{ij})
\partial_+\psi_-^j-\nonumber\\
&&i\partial_j (g_{li}-B_{li})\psi_{+}^i
\partial_- x^l \psi_+^j-i\partial_j (g_{li}+
B_{li})\psi_-^i\partial_+x^l \psi_-^j+
\frac12 R_{ijkl}(\Gamma^{(-)})\psi_+^i 
\psi_+^j\psi_-^k\psi_-^l\}.
\eea
Let us assume that there exists an abelian isometry
$\delta x^i=\epsilon k^i$ (the conditions being the same than
in the pure bosonic case).
In adapted coordinates to the isometry 
$\{\theta,\psi_{\pm}^0,x^\alpha,\psi_{\pm}^\alpha\}$, 
$\alpha=1,\dots,d-1 ,$ 
and retaining only the world-sheet derivative terms of the 
zero components, the 
action reduces to:
\bea
\label{(1.4)}
&&S={1\over2}\int_{\Sigma}d^2\sigma\{g_{00}(\dot\theta^2 - 
\theta^{\prime 2}) + 2j_{-}(\dot\theta + \theta^\prime) + 
2j_{+}(\dot\theta - \theta^\prime) 
-i (g_{00}\psi_{+}^0 + k_{\alpha}^{-}\psi_{+}^{\alpha})
(\dot\psi_{+}^0 - {\psi_{+}^0}^\prime)-\nonumber\\
&&i(g_{00}\psi_-^0+k_{\alpha}^+\psi_-^\alpha)(\dot\psi_{-}^0+
{\psi_-^0}^\prime)+V\}
\eea
where we have defined
\bea
\label{(1.5)}
&&k_{i}^{\pm} =g_{0i} \pm B_{0i}\nonumber\\
&&j_{\pm} ={1\over2}(k_{\alpha}^{\mp}\partial_{\pm}x^\alpha + i
k_{[i,j]}^{\mp}\psi_{\pm}^j\psi_{\pm}^i) \nonumber\\
&&V=(g_{\alpha\beta}+B_{\alpha\beta})\partial_+ x^\alpha
\partial_- x^\beta-i\psi^i_+(g_{i\alpha}+B_{i\alpha})
\partial_-\psi_+^\alpha-i\psi^i_-(g_{i\alpha}-B_{i\alpha})
\partial_+\psi_-^\alpha- \nonumber\\
&&i\partial_j(g_{\alpha i}-B_{\alpha i})
\psi^i_+\partial_-x^\alpha\psi^j_+-i\partial_j
(g_{\alpha i}+B_{\alpha i})\psi^i_-\partial_+x^\alpha
\psi^j_-+\frac12 R^{(-)}_{ijkl}\psi^i_+\psi^j_+\psi^k_-\psi^l_-.
\eea
The canonical momenta associated to the zero coordinates are
\bea
\label{(1.6)}
&&\Pi_{\pm} \equiv  {\delta S\over\delta\dot\psi_{\pm}^0} = 
{i\over2}(g_{00}\psi_{\pm}^0 + k_{\alpha}^{\mp}
\psi_{\pm}^\alpha)\\
\label{(1.61)} 
&&p_{\theta} \equiv {\delta S\over\delta\dot\theta} = 
g_{00}\dot\theta + j_{+} + 
j_{-}
\eea
where (\ref{(1.6)}) are two first class constraints.

The following generating functional:
\be
\label{(1.8)}
{\cal F} = {1\over2}\oint d\sigma\{ \theta^\prime\tilde\theta - 
\theta\tilde\theta^\prime -
i\psi_{+}^0\tilde\psi_{+}^0 +i\psi_{-}^0\tilde\psi_{-}^0 \} 
\ee
yields the abelian T-dual, with backgrounds given
by Buscher's formulas\footnote{$H$
and ${\tilde H}$ are equal up to a spatial derivative
$\frac{i}{2}\partial_\sigma({\tilde \psi}^0_+\psi^0_+
+{\tilde \psi}^0_-\psi^0_-)$. We will see how this term
emerges in the derivation of the generating functional via
the gauging procedure.}.
It induces the change of variables in phase space \cite{H}:
\bea
\label{(1.9)}
&&{\tilde\Pi}_{\pm} = -{\delta F\over\delta\tilde\psi_{\pm}^0} = 
\mp{i\over2}\psi_{\pm}^0, \qquad 
\Pi_{\pm} = {\delta F\over\delta\psi_{\pm}^0} = 
\mp{i\over2}\tilde\psi_{\pm}^0 \nonumber\\
&&p_{\tilde\theta} = -{\delta F\over\delta\tilde\theta} = - 
\theta^\prime ,
\qquad p_{\theta} 
= {\delta F\over\delta\theta} = - \tilde \theta^\prime
\eea 
${\cal F}$ being linear in the 
original and dual variables implies that
the original and dual theories are also
equivalent quantum mechanically, as in the bosonic case.

(\ref{(1.9)}) reads, in configuration space variables:
\bea
\label{(1.9.1)}
&&{\psi}_\pm^0=\mp({\tilde g}_{00}{\tilde \psi}_\pm^0+
{\tilde k}_\alpha^\mp{\tilde \psi}_\pm^\alpha)\nonumber\\
&&\psi_\pm^\alpha={\tilde \psi}_\pm^\alpha
\eea
for the fermions, and:
\bea
\label{(1.9.2)}
&&\partial_+\theta=-{\tilde g}_{00}\partial_+{\tilde \theta}-
{\tilde k}^-_\alpha\partial_+{\tilde x}^\alpha-i
{\tilde k}^-_{[i,j]}{\tilde \psi}^j_+{\tilde \psi}^i_+=
-{\tilde k}^-_i\partial_+{\tilde x}^i-i{\tilde k}^-_{[i,j]}
{\tilde \psi}^j_+{\tilde \psi}^i_+
\nonumber\\
&&\partial_-\theta={\tilde g}_{00}\partial_-{\tilde \theta}+
{\tilde k}^+_\alpha\partial_-{\tilde x}^\alpha+i
{\tilde k}^+_{[i,j]}{\tilde \psi}^j_-{\tilde \psi}^i_- =
{\tilde k}^+_i\partial_-{\tilde x}^i+i{\tilde k}^+_{[i,j]}
{\tilde \psi}^j_-{\tilde \psi}^i_-,
\eea
for the bosons, where 
${\tilde k}^\mp_i={\tilde G}_{0i}\mp {\tilde B}_{0i}$. 
This is the abelian duality mapping in  
the $N=1$ case and it can be obtained as well
from (\ref{141})
replacing bosonic fields by superfields and derivatives by
superderivatives. 

As in the pure bosonic case
(\ref{(1.8)}) can be derived from the total time
derivative term that is induced in the usual gauging procedure. 
Starting with the superspace action (\ref{(1.1)}) we replace the
superderivatives of the adapted coordinate to the isometry by 
covariant superderivatives and add the following 
super-Lagrange multipliers
term \cite{IKR}:
\be
\label{m1}
\int_{\partial\Sigma}(D_+{\tilde X}^0 A_-+D_-{\tilde X}^0 A_+)
\ee
where ${\tilde X}^0$ is the superfield:
${\tilde X}^0={\tilde \theta}+\theta_+{\tilde\psi}_-^0+
\theta_-{\tilde\psi}_+^0+\theta_+\theta_-{\tilde F}$ and $A_\pm$
is the gauge superfield
$A_\pm=f_\pm+\theta_+a^+_\pm+\theta_-a^-_\pm+\theta_+
\theta_-F_\pm$.
The integration on the Lagrange multiplier
imposes the constraint that $A_\pm=D_\pm X^0$, whose substitution
in the gauged action yields the original theory (after fixing the gauge)
plus a total time derivative term. The integration over the gauge fields
$A_\pm$ yields the dual theory. We then have:
\be
\label{m2}
{\tilde S}=S+\int (D_+{\tilde X}^0 D_-X^0+
D_-{\tilde X}^0D_+X^0)
\ee
where $X^0$ is the superfield containing the bosonic coordinate
adapted to the isometry: $X^0=\theta+\theta_+\psi_-^0+\theta_-
\psi_+^0+\theta_+\theta_-F$.
Under a canonical transformation 
$\{q^i,p_i\}\rightarrow\{Q^i,P_i\}$ the generating functional
is such that:
\be
\dot{q}^i p_i-H=\dot{Q}^iP_i-{\tilde H}+\frac{d{\cal F}}{dt}
\ee
where $H$ and ${\tilde H}$ denote the original and canonically
transformed Hamiltonians respectively. 
Hence, from (\ref{m2}) we can read (\ref{(1.8)}) 
after performing the superspace integration.
In this calculation one also obtains the spatial derivative
term that was needed to prove the equivalence between $H$
and ${\tilde H}$.

We now focus on the study of the boundary conditions.
We are going to look at the most general class of 
classical boundary
conditions, i.e. those for which the variation of the 
whole action vanishes. 
Let us consider the general conditions:
\be
\label{i1}
\psi_{+}^i={\bar R}^i(x^k;\psi_{-}^j) 
\ee
for the fermions, where by 
${\bar R}^i(x^k;\psi_-^j)$ we mean:
\be
\label{i11}
{\bar R}^i(x^k;\psi_{-}^j) \equiv \sum_{l={\rm odd}}^d 
{\psi_-^{j_1}...\psi_-^{j_l}\over l!} R_{j_1 ... j_l}^i(x^k),
\ee
i.e. we denote with a bar these kinds of expansions
to distinguish them from ordinary target space 
tensors\footnote{Although
it is clear that usual tensors can be viewed as barred ones 
with vanishing 
expansion except for the zeroth term. With this notation an
arbitrary barred tensor is, in components:
\be
\label{(1.27)}
{\bar V}_{k_1\dots k_m}^{i_1\dots i_n} =
{\bar V}_{k_1\dots k_m}^{i_1\dots i_n}(x;\psi_-) \equiv 
\sum_{l=0}^{d} {\psi_{-}^{j_1}\dots\psi_{-}^{j_l}\over l!} 
V_{k_1\dots k_m j_1\dots j_l}^{i_1\dots i_n}(x).
\ee
}.
The variation of the action in the boundary:
\bea
\label{ii1}
\delta S_{{\rm boundary}}&=&\int dt\{\delta x^i((g_{ij}+B_{ij})
\partial_-x^j-(g_{ij}-B_{ij})\partial_+x^j+i(g_{ij}+B_{ij})_{,k}
\psi^k_-\psi^j_--\nonumber\\
&&i(g_{ij}-B_{ij})_{,k}\psi^k_+\psi^j_+)+i(g_{ij}+B_{ij})
(\delta\psi^i_-\psi^j_--\delta\psi^j_+\psi^i_+)\}
\eea
vanishes if the
following conditions are satisfied:
\be
\label{(1.34)}
(g_{ij}+B_{ij})\p^j =  
(g_{kj}+B_{kj}){\dl{\bar R}^j\over\dl\p^i} {\bar R}^k
\ee
\be
\label{(1.35)}
g_{ij}x^{\prime j}-B_{ij}\dot{x}^j-\frac{i}{2}
(g_{ij}+B_{ij})_{,k}
\p^k\p^j+\frac{i}{2}(g_{ij}-B_{ij})_{,k}
{\bar R}^k {\bar R}^j+\frac{i}{2}(g_{kj}+B_{kj})
{\partial {\bar R}^j\over\partial x^i}{\bar R}^k=0
\ee
at the ends of the string. (\ref{(1.34)}) imposes a constraint
over the possible tensors $R^i_{j_1\dots j_l}$ allowed as
classical boundary conditions.
For instance the linear case: 
$\psi^i_+=J^i\,_j(x)\psi^j_-$ 
considered in \cite{CLNY} is a solution provided $J$ satisfies
$J^t(g+B)J=g-B$. $J$ can be chosen such that $J^t=J$, $J^2=1$
and $BJ=-JB$.

Clearly (\ref{(1.35)}) is not manifestly covariant since
$\partial_i {\bar R}^j$ is not a tensor. 
It can be seen that the covariant derivative of
a general barred tensor (\ref{(1.27)}) is given by:
\be
\label{a4}
{\bar \nabla}_j {\bar V}_{k_1\dots k_m}^{i_1\dots i_n} = 
\nabla_j {\bar V}_{k_1\dots k_m}^{i_1\dots i_n} - \psi_{-}^l 
\Gamma_{jl}^p{\delta{\bar V}_{k_1\dots k_m}^{i_1\dots i_n} 
\over\delta\psi_{-}^p}
\ee
where
\be
\label{a3}
\nabla_j {\bar V}_{k_1\dots k_m}^{i_1\dots i_n} = 
\partial_j {\bar V}_{k_1\dots k_m}^{i_1\dots i_n}
+ \sum_{r=1}^n \Gamma_{ji}^{i_r}
{\bar V}_{k_1\dots k_m}^{i_1\dots i\dots i_n}- 
\sum_{s=1}^m \Gamma_{jk_s}^{k}
{\bar V}_{k_1\dots k\dots k_m}^{i_1\dots i_n},
\ee
since in this way also the target space indices contracted
with the fermions are covariantized.
Then the manifestly covariant bosonic boundary condition is 
given by:
\be
\label{(1.36)}
g_{ij}x^{\prime j}-B_{ij}\dot{x}^j-\frac{i}{2}
(g_{kj}+B_{kj})
{\bar R}^k \ov{\nabla}_i{\bar R}^j -\frac{i}{2} 
\nabla_kB_{ij}(\p^k \p^j + {\bar R}^k {\bar R}^j) =  0
\ee
where use have been made of the constraints (\ref{(1.34)}).
This expression is the generalization to the $N=1$
supersymmetric case of Neumann boundary
conditions.

The linear classical boundary conditions 
$\psi_{+}^i = J^i\,_j(x)\p^j$ are such that the
fermionic boundary contribution to the action:
$B_{ij}(\psi_{+}^i \psi_{+}^j + \p^i \p^j)$ vanishes. This means
that in some sense they are too restrictive since
they do not allow for any fermionic dynamics at the boundary.
We want to investigate if this 
feature is general to all classical boundary conditions.
The variation of
$B_{ij}(\psi_+^i\psi_+^j+\psi_-^i\psi_-^j)$
is given by:
\be
\label{i2}
\frac{\delta}{\delta\psi^k_-}(B_{ij}(\psi^i_-\psi^j_-+
{\bar R}^i{\bar R}^j))
=2(\frac{\delta{\bar R}^j}{\delta\psi^k_-}
{\bar R}^ig_{ij}-g_{kj}\psi^j_-)
\ee
after substitution of the constraints (\ref{(1.34)}).
If we denote 
${\bar E}_k\equiv\frac{\delta}{\delta\psi^k_-}
(B_{ij}(\psi^i_-\psi^j_-+{\bar R}^i{\bar R}^j))$,
the integrability conditions
$\frac{\delta{\bar E}_k}{\delta\psi^l_-}+
\frac{\delta{\bar E}_l}{\delta\psi^k_-}=0$ imply:
\be
\label{i3}
g_{lk}=\frac{\delta{\bar R}^i}{\delta\psi^l_-}
\frac{\delta{\bar R}^j}{\delta\psi^k_-}g_{ij}.
\ee
Substituting in (\ref{i2}) we finally get:
\be
\label{(1.41)}
{1\over2}B_{ij}(\p^i \p^j + {\bar R}^i {\bar R}^j) = 
-g_{ij}{\bar R}^j\frac{\delta{\bar R}^i}{\delta\psi^l_-}
\psi^l_-.
\ee
In the linear case the right-hand side is zero
and therefore the fermionic 
contribution to the boundary action vanishes.
However this is not so in the general case.
The usual R-NS boundary conditions:
$\psi^i_+=\eta\psi^i_-$, with $\eta=\pm 1$,
also produce non-trivial dynamics in the boundary since
they are chosen as minima of the bulk
in (\ref{(1.2)}). They are classical boundary conditions
when $B_{ij}=0$. 

Let us now study how the general classical boundary conditions
(\ref{i1}) and (\ref{(1.36)})
transform under abelian duality.

The conditions for the fermions:
$\psi_{+}^i ={\bar R}^i(x^k;\psi_{-}^j)$,
where ${\bar R}^i$ must have vanishing Lie derivative in
the Killing direction\footnote{This is implied by the
constraints (\ref{(1.34)}).}, transform under 
(\ref{(1.9.1)}) as:
\bea
\label{(1.10)}
&&{\tilde \psi}_{+}^\alpha= 
{\bar R}^{\alpha}({\tilde x}^\gamma;
{\tilde\psi_{-}^0 - k_{\sigma}^{+}
\tilde\psi_{-}^\sigma\over 
g_{00}},\tilde\psi_{-}^\beta)={\tilde R}^\alpha\nonumber\\
&&\tilde\psi_{+}^0 =
-\frac{1}{{\tilde g}_{00}}({\tilde R}^0+{\tilde k}^-_\sigma
{\tilde R}^\sigma),
\eea
where our notation is such that given
a function
${\bar \xi} ={\bar \xi}(x^k;\psi_{-}^0,\psi_{-}^\beta)$ 
${\tilde \xi}$ is defined by: $\tilde\xi \equiv 
\tilde\xi({\tilde x}^k;\tilde\psi_{-}^0,\tilde\psi_{-}^\beta) = 
{\bar \xi}({\tilde x}^k;{\tilde\psi_{-}^0 - 
k_{\sigma}^{+}\tilde\psi_{-}^\sigma\over g_{00}},
\tilde\psi_{-}^\beta)$.  

The corresponding bosonic covariant boundary conditions:
\be
\label{(1.12)}
g_{ij}x^{\prime j} - B_{ij}\dot x^j + 
{\bar W}_i(x^\alpha;\psi_{-}^j) = 0
\ee
with 
\be
\label{ii2}
{\bar W}_i=-\frac{i}{2}(g_{kj}+B_{kj}){\bar R}^k
{\bar \nabla}_i{\bar R}^j-\frac{i}{2}\nabla_k
B_{ij}(\psi^k_-\psi^j_-+{\bar R}^k{\bar R}^j)
\ee
are a little bit more involved.
In phase space variables 
$(\Pi_{\pm},\psi_{\pm}^0,p_{\theta},\theta )$ these 
conditions read:
\be
\label{(1.13)}
g_{i0}\theta^\prime + g_{i\beta}x^{\prime\beta} - 
B_{i0}({p_{\theta} - j_{+} - 
j_{-}\over g_{00}}) - B_{i\beta}\dot x^\beta +{\bar W}_i = 0.
\ee
In dual variables and
splitting the currents into their bosonic and fermionic parts 
$j_{\pm}^f \equiv 
j_{\pm} - {1\over 2}k_{\sigma}^{\mp}\partial_{\pm}x^\sigma$ 
we get for the zero and $\alpha$
components:
\bea
\label{(1.20)}
&&\dot{\tilde\theta} + j_+^f-j_-^f-\tilde W_{0}=0 \\
\label{x1}
&&{\tilde G}_{\alpha i}{\tilde x}^{\prime i} - \tilde B_{\alpha i}
\dot{\tilde x}^{i} + {\tilde G}_{0\alpha}(j_+^f+j_-^f)+
{\tilde B}_{0\alpha}
(j_+^f-j_-^f)+\tilde W_\alpha=0  
\eea
where
\bea
\label{(1.21)}
&&j_+^f=-\frac{i}{2}(\partial_\alpha (\frac{1}{{\tilde k}^2})
{\tilde R}^0 {\tilde R}^\alpha-\partial_\alpha 
(\frac{{\tilde k}_\beta^-}{{\tilde k}^2})
{\tilde R}^\alpha {\tilde R}^\beta)
\nonumber\\
&&j_-^f=-\frac{i}{2}\frac{1}{{\tilde k}^2}(\partial_\alpha
{\tilde k}^2
{\tilde \psi}^\alpha_-{\tilde \psi}^0_-
+\partial_\alpha{\tilde k}^+_\beta
{\tilde\psi}^\alpha_-{\tilde\psi}^\beta_-).
\eea
${\tilde W}_0$, ${\tilde W}_\alpha$ have complicated expressions
in terms of the dual backgrounds and since they will be 
irrelevant in what follows we shall omit them. 
It is clear that for arbitrary backgrounds
(\ref{(1.20)}) cannot be interpreted as a Dirichlet 
boundary condition.
To discuss this further
let us restrict ourselves to the more familiar case of
Neumann R-NS boundary conditions for the initial theory:
\bea
\label{(1.22)}
&&g_{ij}x^{\prime j}-B_{ij}\dot{x}^j=0\nonumber\\
&&\psi^i_+=\eta\psi^i_-;\qquad \eta=\pm 1
\eea
The dual boundary conditions are then:
\bea
\label{(1.23)}
&&{\tilde\psi}^\alpha_+=\eta{\tilde\psi}^\alpha_-\nonumber\\
&&{\tilde\psi}^0_++\eta{\tilde\psi}^0_-=2\eta k_\alpha^*
{\tilde\psi}^\alpha_-\nonumber\\
&&\dot{\tilde\theta}=i\partial_\alpha k_\beta^*
{\tilde\psi}^\alpha_-
{\tilde\psi}^\beta_-\nonumber\\
&&{\tilde G}_{\alpha i}{\tilde x}^{\prime i}-{\tilde B}_{\alpha i}
{\dot{\tilde x}}^i=-ik_\alpha^*\partial_\beta{\tilde k}^2
{\tilde\psi}^\beta_-
{\tilde\psi}^0_-+i({\tilde k}^\alpha_+\partial_\beta
k^*_\sigma-k^*_\alpha
\partial_\beta
{\tilde k}^+_\sigma){\tilde\psi}^\beta_-{\tilde\psi}^\sigma_-
\eea
where $k^*_\alpha=B_{0\alpha}$. These results are in agreement 
with (3.12) in
\cite{ABB}.
The non-trivial terms (those that spoil Dirichlet NS-R boundary
conditions in the dual) are all proportional to $B_{0\alpha}$.
Therefore a super D-brane is obtained in the dual only if the
original background is such that $B_{0\alpha}=0$.
If this occurs
(\ref{(1.23)}) turns into: 
${\tilde\psi}^0_+=-\eta{\tilde\psi}^0_-$, accounting for the
reversal of space-time chirality under T-duality
\cite{DHS},
$\dot{\tilde\theta}=0$, and Neumann R-NS boundary 
conditions for the rest of the coordinates. 
This is the case, in particular, for the type I superstring
where the D-brane is actually an orientifold.
In this theory consistency conditions restrict the possible
D-manifolds to one, five and nine-branes \cite{PCJ}.
Since the only consistent open superstring theory is the
type I superstring, which contains unoriented topologies, it
is interesting to analyze in a little bit more detail the
unoriented world-sheets. As in
the previous section we consider the projective plane,
obtained from the disk by identifying opposite points.
Crosscap boundary conditions for the fermions contain 
an $i$ factor
due to the fact that we are taking a constant time boundary
\cite{CLNY}:
\bea
\label{cr1}
&&\psi_{+}^i (\sigma + \pi) = i\eta \p^i (\sigma),\qquad 
\eta=\pm 1 \\
\label{cr2}
&&x^{\prime i}(\sigma + \pi) = x^{\prime i}(\sigma)\nonumber\\
&&\dot x^i(\sigma + \pi) = -\dot x^i(\sigma).
\eea
These conditions 
are mapped under (\ref{(1.9.1)}) and (\ref{(1.9.2)}) to:
\bea
\label{cr3}
&&\tilde \psi_{+}^\alpha (\sigma + \pi) = 
i\eta\tilde \p^\alpha (\sigma) \nonumber\\
&&\tilde\psi_{+}^0 (\sigma + \pi) = -i\eta \tilde\p^0 (\sigma)
\eea
for the fermions, giving the usual change of sector for the
0-component, and to:
\bea
\label{(1.28)}
&&\dot{\tilde\theta}(\sigma + \pi) = \dot{\tilde\theta}(\sigma)
\nonumber\\
&&\tilde{\theta^\prime}(\sigma + \pi) = 
-\tilde{\theta^\prime}(\sigma) \nonumber\\
&&\dot{{\tilde x}}^\alpha(\sigma+\pi)
=-\dot{{\tilde x}}^\alpha(\sigma)\nonumber\\
&&{\tilde x}^{\prime\alpha}(\sigma+\pi)=
{\tilde x}^{\prime\alpha}(\sigma)
\eea
for the bosons, i.e. orientifold conditions for the 
${\tilde \theta}$ coordinate and crosscap for the rest. 
Therefore the dual theory is an orientifold, static since the abelian 
electromagnetic field is absent for unoriented strings.

\section{Non-abelian duality}
\setcounter{equation}{0}

In this section we analyze non-abelian duality transformations
in backgrounds of open and closed strings. 
We follow the canonical transformations approach. We argue that
this approch is more suitable to discuss non-abelian duality
in world-sheets with non-trivial genera or boundaries, therefore for
open strings, since it doesn't convey some subtleties involved in the
more usual gauging procedure \cite{DLOQ}. 
The gauging approach has been applied to backgrounds
of open and closed strings in \cite{FKS}. 
Also, particular subclasses of the backgrounds we will be considering
have been studied in \cite{KS} within the Poisson-Lie T-duality.

Let us consider a sigma model:
\be
\label{2uno}
S=\int_\Sigma d\sigma_+d\sigma_-
(g_{ij}+B_{ij})\partial_+x^i\partial_-x^j
\ee
where $B_{ij}=b_{ij}+F_{ij}$, with a non-abelian
isometry group $G$ generated by the 
Killing vectors $k^i_a, a=1,\dots,{\rm dim}G$. In order to construct
the non-abelian dual with respect to this set of isometries following
the gauging procedure \cite{DLOQ} we
need to have invariance
under the local gauge transformations 
$\delta x^i=\epsilon^a (\sigma) k^i_a(x)$.
This is accomplished by introducing 
gauge fields in the Lie algebra $A^a_{\pm}$,
with gauge variation
$\delta A^a_\pm=-\partial_\pm\epsilon^a+
f^a_{bc}A^b_\pm \epsilon^c$ ($f_{abc}$ are the structure 
constants of the Lie algebra), and replacing
ordinary derivatives by covariant derivatives:
\be
\label{2tres}
\partial_\pm x^i\rightarrow D_\pm x^i=\partial_\pm x^i+k^i_a A^a_\pm.
\ee
The gauged action reads:
\be
\label{2cinco}
S_{{\rm gauged}}=S+\int d\sigma_+ d\sigma_- (g_{ij}+B_{ij})
(k^i_ak^j_bA^a_+A^b_-+A^a_-k^j_a\partial_+x^i+A^a_+k^i_a\partial_-x^j).
\ee
In homologically trivial world-sheets the
Lagrange mutipliers term
\be
\label{2seis}
-\int_\Sigma Tr (\chi(dA+A\wedge A))
\ee
is added
to impose the pure gauge condition $A=-dgg^{-1}$,
$g\in G$, upon integration on $\chi$ and thus recover the original
theory after fixing the gauge.
For higher genus or in the presence of boundaries
it is not known how to impose the pure gauge
constraint in a gauge invariant way. 
This problem already appeared
in closed strings when trying to generalize \cite{DLOQ}
to arbitrary genera \cite{AABL,GR}. In particular, 
when homologically non-trivial cycles exist
in the world-sheet 
the integration on $\chi$ in (\ref{2seis})
implies:
\be
\label{2seis1}
A=-d(gg_c)(gg_c)^{-1}
\ee
where we have the contribution of the non-trivial flat conexion along the
boundaries and the $2h$ cycles ($h$ is the genus of the world-sheet):
\be
\label{2seis2}
g_c=P e^{\oint A}
\ee
(here $P$ stands for path ordering).
Imposing $g_c=1$ in a gauge invariant way introduces non-local terms
in the action, 
spoiling the non-abelian dual construction.
In closed strings it is
possible to deal with higher genera using the canonical transformation
approach \cite{CZ,L,AL,CS1}. The idea is that once we have a relation 
like (\ref{1cinco1}) between the
Hilbert spaces of the original and dual theories, 
$|\theta(\sigma)\rangle$ can be the result of integrating the
original theory on an arbitrary Riemann surface with boundary.
This means in particular that we can apply the canonical
description to open string world-sheets. Non-abelian duality
as an explicit canonical transformation
is only known in those
cases in which the isometry acts without isotropy, 
i.e. without fixed
points, therefore restricting the backgrounds 
where non-abelian
duality can be studied. The most general sigma model of this 
kind is \cite{GR}:
\bea
\label{2once}
S[g,x]=&&\int d\sigma_+d\sigma_-[E_{ab}(x)(\partial_+g 
g^{-1})^a (\partial_-g g^{-1})^b+F^R_{a\alpha}(x)(\partial_+g g^{-1})^a
\partial_- x^\alpha+ \nonumber\\
&&F^L_{\alpha a}(x)\partial_+ x^\alpha 
(\partial_- g g^{-1})^a+F_{\alpha\beta}(x)\partial_+ x^\alpha
\partial_- x^\beta],
\eea
where $g\in G$, a Lie group (which we take to be compact), and
$\partial_\pm g g^{-1}=(\partial_\pm g g^{-1})^a T_a$ with $T_a$ the
generators of the corresponding Lie algebra\footnote{$\{T_a\}$
are normalized such that $Tr(T_aT_b)=\delta_{ab}$.}.
This model is invariant under $g\rightarrow gh$, with $h\in G$. 
We can regard  
(\ref{2once}) as a background of open and closed strings (neutral case)
with the abelian gauge field absorbed in the corresponding
torsion terms. 
  
Let us parametrize the
Lie group using the Maurer-Cartan forms $\Omega^a_k$, such that
\be
\label{2doce}
(\partial_\pm g g^{-1})^a=\Omega^a_k(\theta) \partial_\pm\theta^k.
\ee
The abelian background gauge fields that are compatible with the
non-abelian isometry $g\rightarrow gh$ have the form
$V_i=\frac12 \Omega^a_iC^a(x)$, with $C$ arbitrary, 
and $V_\alpha$ $\theta^i$-independent.
Then $E_{[ab]}=b_{ab}+f_{abc}C^c(x)$, where $b_{ab}$ is the closed
strings antisymmetric tensor and
$F^R_{a\alpha}=f^R_{a\alpha}-\frac12\partial_\alpha C^a(x)$,
$F^L_{\alpha a}=f^L_{\alpha a}+\frac12\partial_\alpha C^a(x)$
(with $f^R$, $f^L$ the corresponding closed
strings backgrounds).
 
The following canonical transformation from
$\{\theta^i,\Pi_i\}$ to $\{\chi^a,{\tilde \Pi}_a\}$:
\bea
\label{2trece}
&&\Pi_i=-(\Omega^a_i {\chi^\prime}^a+f_{abc}\chi^a
\Omega^b_j\Omega^c_i{\theta^\prime}^j) \nonumber\\
&&{\tilde \Pi}_a=-\Omega^a_i{\theta^\prime}^i
\eea
produces the non-abelian dual of (\ref{2once}) with respect to its
isometry $g\rightarrow gh$:
\be
\label{2catorce}
{\tilde S}=\int d\sigma_+ d\sigma_-
[(E+{\rm ad}\chi)^{-1}_{ab}(\partial_+\chi^a+F^L_{\alpha a}(x)
\partial_+ x^\alpha)(\partial_-\chi^b-F^R_{b\beta}(x)
\partial_- x^\beta)+F_{\alpha\beta}\partial_+x^\alpha
\partial_-x^\beta]
\ee
This was first realized in \cite{CZ} for the case of $SU(2)$ principal
chiral models (where $E_{ab}=\delta_{ab}$, $F^R_{a\alpha}=
F^L_{\alpha a}=F_{\alpha\beta}=0$), generalized in 
\cite{L,AL} to arbitrary group, and shown to apply also 
to this more general case in \cite{S}. 
The dual backgrounds are given by:
\bea
\label{2quince1}
&&{\tilde G}_{ab}=\frac12 M_{(ab)}, \qquad
{\tilde B}_{ab}=\frac12 M_{[ab]}, \qquad
{\tilde G}_{a\alpha}=\frac12 (-M_{ab}F^R_{b\alpha}+M_{ba}F^L_{\alpha b})
\nonumber\\
&&{\tilde B}_{a\alpha}=-\frac12 (M_{ab}F^R_{b\alpha}+
M_{ba}F^L_{\alpha b}), \qquad
{\tilde G}_{\alpha\beta}=-\frac12 M_{ab}(F^L_{\alpha a}F^R_{b\beta}+
F^L_{\beta a}F^R_{b\alpha})+\frac12 F_{(\alpha\beta)} \nonumber\\
&&{\tilde B}_{\alpha\beta}=-\frac12 M_{ab}(F^L_{\alpha a}F^R_{b\beta}-
F^L_{\beta a}F^R_{b\alpha})+\frac12 F_{[\alpha\beta]}
\eea
where we use capital letters for the metric and torsion to account
for the fact that the initial abelian background gauge field
is absorbed in $M$ and $F^L, F^R, F$.
$M\equiv (E+{\rm ad}\chi)^{-1}$, $({\rm ad}\chi)_{ab}=
f_{abc}\chi^c$ and
$M_{(ab)}=M_{ab}+M_{ba}$, $M_{[ab]}=M_{ab}-M_{ba}$.

(\ref{2trece}) reads, in configuration space variables:
\bea
\label{e1}
&&\Omega^a_i\partial_+\theta^i=-M_{ba}(\partial_+\chi^b+
F^L_{\alpha b}\partial_+ x^\alpha)=
-({\tilde G}_{ab}-{\tilde B}_{ab})\partial_+\chi^b-
({\tilde G}_{a\alpha}-{\tilde B}_{a\alpha})
\partial_+x^\alpha \nonumber\\
&&\Omega^a_i\partial_-\theta^i=M_{ab}(\partial_-\chi^b-
F^R_{b\alpha}\partial_-x^\alpha)=
({\tilde G}_{ab}+{\tilde B}_{ab})\partial_-\chi^b+
({\tilde G}_{a\alpha}+{\tilde B}_{a\alpha})
\partial_-x^\alpha.
\eea
These relations generalize (\ref{141}) for
non-abelian duality transformations, the main difference being that
the components of the torsion in the Lie algebra variables
appear explicitly. They identify the  
non-local transformation responsible for non-abelian
duality.

Up to now only the transformations in the bulk have been accounted for.
In order to
study the open strings sector we have to analyze as well the boundary
conditions.
Using (\ref{e1}) it is easy to see that Neumann boundary conditions
in the initial theory:
\be
\label{2dieciocho}
E_{(ab)}\Omega^b_j{\theta^\prime}^j-E_{[ab]}\Omega^b_j{\dot\theta}^j
-F^R_{a\alpha}\partial_-x^\alpha+F^L_{\alpha a}\partial_+x^\alpha
=0
\ee
imply for the dual:
\be
\label{2diecinueve}
(2\delta_{ab}-({\rm ad}\chi)_{ac} M_{[cb]}){\dot\chi}^b
+({\rm ad}\chi)_{ab}
(M_{(bc)}\chi^{\prime c}+M_{bc}F^R_{c\alpha}\partial_-x^\alpha+
M_{cb}F^L_{\alpha c}\partial_+x^\alpha)=0
\ee
or
\bea
\label{2.19.1}
&&E_{(ab)}({\tilde G}_{bc}\dot{\chi}^c-{\tilde B}_{bc}
\chi^{\prime c}+{\tilde G}_{b\alpha}\dot{x}^\alpha-
{\tilde B}_{b\alpha}x^{\prime\alpha})-E_{[ab]}
({\tilde G}_{bc}\chi^{\prime c}-{\tilde B}_{bc}\dot{\chi}^c
+{\tilde G}_{b\alpha}x^{\prime\alpha}-{\tilde B}_{b\alpha}
\dot{x}^\alpha)- \nonumber\\
&&(F^L_{\alpha a}-F^R_{a\alpha})\dot{x}^\alpha
-(F^L_{\alpha a}+F^R_{a\alpha})x^{\prime\alpha}=0,
\eea
in terms of the dual backgrounds. Therefore the general dual
boundary conditions are a combination of Dirichlet and
Neumann boundary conditions with some additional terms depending
on the inert coordinates.
For the $x^\alpha$ coordinates it can be seen that
\be
\label{2veintidos1}
g_{\alpha i}{\theta^\prime}^i+g_{\alpha\beta}x^{\prime\beta}
-B_{\alpha i}{\dot\theta}^i-B_{\alpha\beta}{\dot{x}}^\beta=0
\ee
implies:
\be
\label{2veintidos2}
{\tilde G}_{\alpha a}\chi^{\prime a}+{\tilde G}_{\alpha\beta}
x^{\prime\beta}-{\tilde B}_{\alpha a}{\dot\chi}^a-
{\tilde B}_{\alpha\beta}{\dot{x}}^\beta=0,
\ee
i.e. they still verify Neumann boundary conditions.

Let us now study those cases in which (\ref{2.19.1}) reduces
to generalized Dirichlet conditions. 
First let us make the change of variables:
\be
\label{uu2}
\chi^a\equiv \chi^a+\frac12 C^a(x)
\ee
in order to eliminate the
dependence of the matrix $M$ on the background gauge field.
Dirichlet conditions are obtained when the initial torsion 
$E_{[ab]}=b_{ab}+f_{abc}C^c$
is zero and also
$F^R_{a\alpha}=F^L_{\alpha a}=0$. 
Then (\ref{2.19.1}) reduces to:
\be
\label{uu1}
{\tilde g}_{ab}\dot{\chi}^b-{\tilde b}_{ab}{\chi}^{\prime b}
=0
\ee
where ${\tilde g}, {\tilde b}$ are the closed strings 
backgrounds (given by (\ref{2quince1}) without the contribution
of the background gauge field). 
We can then conclude that
for certain kinds of sigma-models with non-abelian isometries a 
curved $(d-{\rm dim}G-1)$ D-brane is obtained in the dual with 
metric ${\tilde g}_{ab}$ and torsion ${\tilde b}_{ab}$. In this
more general case it is not the velocity that vanishes at the ends
of the string but the whole momentum associated to the 
non-flat background.
The transverse dynamics of the brane would be
given by (\ref{uu2}), therefore induced by the background gauge 
fields: 
$\chi^a\equiv\chi^a+\frac12 C^a(x)=\chi^a+\omega^{ai}V_i$,
with $\omega^{ai}$ such that 
$\omega^{ai}\Omega^b_i=\delta^{ab}$,
but the condition $E_{[ab]}=0$ implies that the 
$i$-components
of the antisymmetric
tensor and the background gauge field must be zero
and therefore the brane is static.
The backgrounds of unoriented strings are among
the ones for which we get Dirichlet boundary conditions in the
dual. We will analyze in the next subsection their $N=1$
supersymmetrization.

It is important to point out that the conditions that must be satisfied
in order to obtain a Dirichlet 
brane in the dual ($E_{[ab]}=F^L_{\alpha a}=F^R_{a\alpha}=0$) are
precisely those required to have a symmetry under 
$g\rightarrow hg$, $h\in G$, in the initial 
sigma-model\footnote{We need to have as well $E_{(ab)}=
f(x^\alpha)\delta_{ab}$ but this can always be achieved by choosing
a proper normalization for the generators of the Lie algebra.}.
In these cases we also find a symmetry in the dual theory under
$\chi$ transforming in the adjoint representation\footnote{This 
symmetry was expected since
the left and right symmetries of the original theory commute
and we are dualizing with respect to the right action only,
so we should get a reminiscent of the left symmetry in the dual.}
and it is easy to see that the canonical transformation couples
the conserved currents associated to the left
symmetry of the initial theory:
\bea
\label{uu3}
&&J^{a(L)}_+=\frac12 E_{(ab)}\Omega^b_i\partial_+\theta^i\nonumber\\
&&J^{a(L)}_-=\frac12 E_{(ab)}\Omega^b_i\partial_-\theta^i
\eea
and the ones associated to $\chi\rightarrow h\chi h^{-1}$ in the
dual\footnote{Up to a total derivative term which for 
principal chiral models ($E_{ab}=\delta_{ab}$) is the responsible
for having curvature free currents in the dual, that are coupled
to the curvature free currents of the principal chiral model
\cite{CZ,L}.}:
\bea
\label{uu4}
&&{\tilde J}^a_+=\partial_+\chi^a-\frac12 E_{(ab)}M_{cb}
\partial_+\chi^c\nonumber\\
&&{\tilde J}^a_-=-\partial_-\chi^a+\frac12 E_{(ab)}M_{bc}
\partial_-\chi^c.
\eea
The existence of a symmetry in the dual theory allows to
establish the equivalence between the original and dual theories
at the level of the Hilbert spaces.
The generating functional responsible for the canonical 
transformation (\ref{2trece}) is:
\be
\label{2quince}
{\cal F}[\chi, \theta]=\oint d\sigma Tr(\chi\partial_\sigma
g g^{-1})
\ee
i.e. it is linear in the dual variables but non-linear in
the original ones. This means that in general it will receive
quantum corrections when implemented at the level of the Hilbert
spaces \cite{GC}, the reason being that we cannot prove a relation
like
\be
\label{e2}
|\chi^a\rangle=\int {\cal D}\theta^i(\sigma)
e^{i{\cal F}[\chi^a,\theta^i(\sigma)]}|\theta^i(\sigma)\rangle
\ee
using the eigenfunctions of the respective Hamiltonians.
However, it was shown in \cite{CZ,L} that such a relation
can in fact be proven using the eigenfunctions of the respective
conserved currents in the initial and dual theories. 
Of course
for this to be true we need to have a symmetry in the dual
theory, which is not the case for arbitrary backgrounds.
As in the abelian case there can still be renormalization effects
modifying the classical backgrounds.
We should mention here that a dilaton shift is needed in order
to preserve conformal invariance \cite{DLOQ}, exactly as in
the abelian case. This remains an open question within the
canonical transformation description whose resolution we
believe should be along the lines previously mentioned in the
abelian case.

Therefore we can conclude that the non-abelian dual of the
subclass of sigma-models (\ref{2once}) 
for which the canonical transformation description
is valid quantum mechanically (up to the dilaton problem and
the renormalization effects we
have just mentioned) is a curved $(d-{\rm dim}G-1)$ D-brane.
More general cases have duals that cannot be interpreted as 
Dirichlet branes. 

Let us finish this section with a few comments on unoriented strings. 
The previous backgrounds 
($E_{[ab]}=F^L_{\alpha a}=F^R_{a\alpha}=0$)
are further restricted by
$F_{\alpha\beta}=\frac12 F_{(\alpha\beta)}$.
Crosscap conditions for the
$\theta^i$-coordinates:
\bea
\label{2veintitres}
&&\dot{\theta}^i(\sigma+\pi)=-\dot{\theta}^i(\sigma)
\nonumber\\
&&{\theta^\prime}^i(\sigma+\pi)={\theta^\prime}^i(\sigma)
\eea
are mapped under (\ref{e1}) to
\bea
\label{2veinticuatro}
&&{\tilde g}_{ab}\chi^{\prime b}(\sigma+\pi)
-{\tilde b}_{ab}\dot{\chi}^b(\sigma+\pi)=-
({\tilde g}_{ab}\chi^{\prime b}(\sigma)-{\tilde b}_{ab}
\dot{\chi}^b(\sigma)) \nonumber\\
&&{\tilde g}_{ab}\dot{\chi}^b(\sigma+\pi)
-{\tilde b}_{ab}\chi^{\prime b}(\sigma+\pi)=
{\tilde g}_{ab}\dot{\chi}^b(\sigma)-{\tilde b}_{ab}
\chi^{\prime b}(\sigma).
\eea
Thus we find
a curved orientifold in the dual with metric and torsion
${\tilde g}_{ab}, {\tilde b}_{ab}$ and generalized
orientifold conditions given by equal momenta at the
identification points (the second equation
in (\ref{2veinticuatro})) and opposite sign momentum
flows out of them.
This is the generalization to curved orientifolds of
the usual conditions:
\bea
&&\dot{x}^i(\sigma+\pi)=\dot{x}^i(\sigma)\nonumber\\
&&x^{\prime i}(\sigma+\pi)=-x^{\prime i}(\sigma).
\eea

\subsection{Superstrings}

Let us now study the $N=1$ supersymmetrization of 
the non-abelian models considered.
Non-abelian duality in these theories
has been studied in \cite{T,S} and more extensively
in \cite{CZ2,CZ3} for principal chiral models. Our aim here is to
focus on the mapping of the boundary conditions. For simplicity we
are going to restrict ourselves to the case of principal
chiral models. 
Their $N=1$ supersymmetrization is given by (\ref{(1.2)})
with $g_{ij}=\Omega^a_i\Omega^a_j$ and $b_{ij}=0$. We take as
well zero background gauge fields, so that the examples we
will be considering are suitable as backgrounds of 
unoriented strings, which are the ones interesting to us
since the
only consistent open superstring theory, the type I superstring,
contains unoriented strings.

Following \cite{CZ3} we can use tangent space variables 
for the fermions 
$\phi^a_\pm=\Omega^a_i \psi^i_\pm$, in which case the action
reads:
\bea
\label{u1}
S&=&\int_\Sigma d\sigma_+ d\sigma_-[(\partial_+gg^{-1})^a
(\partial_-gg^{-1})^a-i\phi^a_+\partial_-\phi^a_+
-i\phi^a_-\partial_+\phi^a_-+\frac{i}{2}f_{abc}
\phi^a_+(\partial_-gg^{-1})^b\phi^c_++\nonumber\\
&&\frac{i}{2}f_{abc}\phi^a_-(\partial_+gg^{-1})^b\phi^c_-+
\frac18 f_{adb}f_{bce}\phi^a_+\phi^d_+\phi^c_-\phi^e_-].
\eea
Working in phase space variables $\{(\theta^i,\Pi_i),
(\phi^a_\pm,\Pi^a_{\phi\pm})\}$:
\bea
\label{u2}
&&\Pi_i=\Omega^a_i\Omega^a_j{\dot\theta}^j
+\frac{i}{4} f_{abc}
\Omega^b_i (\phi^a_+\phi^c_++\phi^a_-\phi^c_-)\\
\label{u3}
&&\Pi^a_{\phi\pm}=\frac{i}{2} \phi^a_\pm,
\eea
where
(\ref{u3}) are a set of first class constraints, the non-abelian
dual of (\ref{u1}) with respect to the right action of the whole
symmetry group $G$ can be obtained through a canonical 
transformation from 
$\{(\theta^i,\Pi_i)\}, (\phi^a_\pm,\Pi^a_{\phi\pm})\}$
to $\{(\chi^a,{\tilde \Pi}_a), ({\tilde \phi}^a_\pm, 
{\tilde \Pi}^a_{{\tilde \phi}\pm})\}$. Namely:
\bea
\label{u4}
&&\Pi_i=-(\Omega^a_i\chi^{\prime a}+f_{abc}\chi^a\Omega^b_j
\Omega^c_i\theta^{\prime j}) \nonumber\\
&&{\tilde \Pi}_a=-(\Omega^a_i \theta^{\prime i}+\frac{i}{4}
f_{abc}(\phi^b_+\phi^c_+-\phi^b_-\phi^c_-))
\eea
for the bosonic momenta, and:
\bea
\label{u5}
&&\Pi^a_{\phi\pm}=\mp\frac{i}{2}({\tilde \phi}^a_\pm+f_{abc}\chi^b
\phi^c_\pm) \nonumber\\
&&{\tilde \Pi}^a_{{\tilde \phi}\pm}=\mp \frac{i}{2}\phi^a_\pm
\eea
for the fermionic ones. 
Its generating functional is:
\be
\label{u6}
{\cal F}=\oint d\sigma [\chi^a \Omega^a_i\theta^{\prime i}
+\frac{i}{4}
f_{abc}\chi^a(\phi^b_+\phi^c_+-\phi^b_-\phi^c_-)-\frac{i}{2}
(\phi^a_+{\tilde \phi}^a_+-\phi^a_-{\tilde \phi}^a_-)].
\ee
As in the abelian case $H$ and ${\tilde H}$ are related up to
a spatial derivative, which in this case is:
\be
\label{u99}
\frac{i}{2}\partial_\sigma({\tilde \phi}^a_-\phi^a_-+
{\tilde \phi}^a_+\phi^a_+ +\frac12 ({\rm ad}\chi)_{ab}
(\phi^a_-\phi^b_- +\phi^a_+\phi^b_+)).
\ee
The dual action is given by:
\bea
\label{u7}
{\tilde S}&=&\int_{\Sigma}d\sigma_+ d\sigma_-[M_{ab}(\partial_+
\chi^a\partial_-\chi^b-i{\tilde \phi}^a_+\partial_-
{\tilde \phi}^b_++i\partial_+{\tilde \phi}^a_-
{\tilde \phi}^b_-)+ iM_{bc}f_{cde}M_{da}{\tilde \phi}^a_-
{\tilde \phi}^e_-\partial_+\chi^b+\nonumber\\
&&iM_{ac}f_{cde}M_{db}{\tilde \phi}^a_+{\tilde \phi}^e_+
\partial_-\chi^b+
L_{abcd}{\tilde \phi}^a_+{\tilde \phi}^b_-{\tilde \phi}^c_+
{\tilde \phi}^d_-] 
\eea
with $M=(1+{\rm ad}\chi)^{-1}$ and
$L_{abcd}=-(f_{agf}f_{ieb}+f_{aie}f_{gfb})M_{ci}M_{eg}M_{fd}$.
(\ref{u7}) is manifestly $N=1$ supersymmetric having the
form of (\ref{(1.3.1)}) with variables living in the Lie algebra
and metric and antisymmetric tensor $\frac12 M_{(ab)}$,
$\frac12 M_{[ab]}$ respectively.
The dual momenta are:
\bea
\label{u8}
&&{\tilde \Pi}_a=\frac12 (M_{(ab)}{\dot\chi}^b-M_{[ab]}
\chi^{\prime b}
-i(M{\rm ad}{\tilde \phi}_-M)_{ab}{\tilde \phi}^b_-
-i(M{\rm ad}{\tilde \phi}_+M)_{ba}{\tilde \phi}^b_+) \\
\label{u88}
&&{\tilde \Pi}^a_{{\tilde \phi}+}=\frac{i}{2}M_{ba}{\tilde \phi}^b_+
\nonumber\\
&&{\tilde \Pi}^a_{{\tilde \phi}-}=\frac{i}{2}M_{ab}{\tilde \phi}^b_-
\eea
where $({\rm ad}{\tilde \phi}_\pm)_{ab}=f_{abc}{\tilde \phi}^c_\pm$.
{}From (\ref{u88}) and (\ref{u5}) we see that the fermions simply 
transform with the change
of scale:
\bea
\label{u9}
&&\phi^a_+=-M_{ba}{\tilde \phi}^b_+ \nonumber\\
&&\phi^a_-=M_{ab}{\tilde \phi}^b_-.
\eea
The corresponding non-local transformation for the bosonic
part is given by:
\bea
\label{mm1}
&&\Omega^a_i\partial_+\theta^i=-M_{ba}\partial_+\chi^b+
\frac{i}{2}(f_{dab}M_{fb}M_{ed}+2f_{cde}M_{da}M_{fc})
{\tilde \phi}^e_+{\tilde \phi}^f_+\nonumber\\
&&\Omega^a_i\partial_-\theta^i=M_{ab}\partial_-\chi^b+
\frac{i}{2}(f_{dab}M_{de}M_{bf}+2f_{cde}M_{ad}M_{cf})
{\tilde \phi}^e_-{\tilde \phi}^f_-.
\eea
In terms of the dual backgrounds this corresponds to:
\bea
\label{mm2}
&&(\partial_+g g^{-1})^a=-({\tilde g}_{ab}-{\tilde b}_{ab})
\partial_+\chi^b-i\partial_e({\tilde g}_{ab}-
{\tilde b}_{ab}){\tilde \phi}^e_+{\tilde \phi}^b_+
-i(\phi_+^2)^a \nonumber\\
&&(\partial_-g g^{-1})^a=({\tilde g}_{ab}+{\tilde b}_{ab})
\partial_-\chi^b+i\partial_e({\tilde g}_{ab}+{\tilde b}_{ab})
{\tilde \phi}^e_-{\tilde \phi}^b_--i(\phi^2_-)^a,
\eea
where the last terms need still be written in terms of the
dual fermions. In this form we see that they generalize 
(\ref{(1.9.2)}) by means of the last quadratic terms in the
fermions, which are zero in the abelian case.
(\ref{u9}) and (\ref{mm1}) can also be obtained from 
the corresponding
(\ref{e1}) in superspace \cite{S} by introducing chiral
superfields \cite{CZ3}.
As in the pure bosonic non-abelian
case the canonical transformation couples the conserved currents
associated to the left symmetry $g\rightarrow hg$ of the initial
theory and the ones associated to transformations in the adjoint in
the dual. Namely:
\bea
\label{nn1}
J^{a(L)}_+&=&(\partial_+ gg^{-1})^a+i(\phi^2_+)^a=
\Omega^a_i\partial_+\theta^i+\frac{i}{2}
f_{abc}\phi^b_+\phi^c_+ \nonumber\\
J^{a(L)}_-&=&(\partial_- gg^{-1})^a+i(\phi^2_-)^a=
\Omega^a_i\partial_-\theta^i+\frac{i}{2}
f_{abc}\phi^b_-\phi^c_-
\eea
with
\bea
\label{nn2}
{\tilde J}^a_+&=&\partial_+\chi^a-M_{ba}\partial_+\chi^b+
i(M{\rm ad}{\tilde \phi}_+M)_{ba}{\tilde \phi}^b_+
\nonumber\\
{\tilde J}^a_-&=&-\partial_-\chi^a+M_{ab}\partial_-\chi^b-
i(M{\rm ad}{\tilde \phi}_-M)_{ab}{\tilde \phi}^b_-.
\eea
Then we can also establish the quantum equivalence between the two
theories, since it is easy to check that the following
relation holds\footnote{Up to the same total derivative of the
bosonic case.}:
\be
\label{nn3}
{\tilde J}^a_\pm e^{iF}=J^{a(L)}_\pm e^{iF}.
\ee
We can then prove
\be
\label{nn4}
{\tilde \psi}_k[\chi,{\tilde \phi}]=N(k)\int
\prod_{i=1}^{{\rm dim}G}{\cal D}\theta^i(\sigma)
e^{iF[\chi,{\tilde \phi},\theta(\sigma),\phi]}
\psi_k[\theta(\sigma),\phi]
\ee
with ${\cal F}$ given by the classical expression (\ref{u6}) 
and $\psi_k$ and ${\tilde \psi}_k$ eigenfunctions of
the respective conserved currents with the same eigenvalue.

Let us now study the mapping of the boundary conditions. 
For simplicity we are going to restrict ourselves to R-NS
boundary conditions for the fermions\footnote{In these 
models they are classical 
boundary conditions since we have null torsion.}:
\be
\label{u10}
\phi^a_+=\eta\phi^a_-,\qquad \eta=\pm 1.
\ee
{}From (\ref{u9}) we get in the dual:
\be
\label{u11}
{\tilde \phi}^a_+=-\eta M_{ba}^{-1}M_{bc}{\tilde \phi}^c_-,
\ee
which are not NS-R boundary conditions. We can just point out
that they could be interpreted as NS-R plus corrections in
${\rm ad}\chi$.

Concerning the bosons, if we start with Neumann boundary conditions:
$\Omega^a_i\theta^{\prime i}=0$, we get in the dual:
\be
\label{u12}
M_{(ab)}{\dot\chi}^b-M_{[ab]}\chi^{\prime b}-i(M_{ac}f_{cde}M_{db}
-\frac12 f_{adc}M_{db}M_{ce}){\tilde \phi}^e_-{\tilde \phi}^b_-
-i(M_{bc}f_{cde}M_{da}+\frac12 f_{adc}M_{bd}M_{ec})
{\tilde \phi}^e_+{\tilde \phi}^b_+=0,
\ee
which in terms of the dual backgrounds can be written as:
\be
\label{u13}
{\tilde g}_{ab}\dot{\chi}^b-{\tilde b}_{ab}\chi^{\prime b}+
\frac{i}{2}\partial_e({\tilde g}_{ab}+{\tilde b}_{ab})
{\tilde \phi}^e_-{\tilde \phi}^b_-+\frac{i}{2}\partial_e
({\tilde g}_{ab}-{\tilde b}_{ab}){\tilde \phi}^e_+
{\tilde \phi}^b_+=0.
\ee
This equation represents the vanishing of ${\tilde \Pi}_a$
(given by (\ref{u8}))
at the ends of the string. Therefore we find a curved
$N=1$ supersymmetric D-brane
(in this particular example (-1)-brane, since we haven't 
allowed
for inert coordinates) with metric and torsion given by
${\tilde g}_{ab}$ and ${\tilde b}_{ab}$.
However since the only consistent open superstring theory 
contains unoriented topologies the D-brane is an orientifold,
as happened in the abelian case. In particular, one can see
that crosscap boundary conditions are mapped to:
\bea
\label{z2}
&&{\tilde \Pi}_a(\sigma+\pi)={\tilde \Pi}_a(\sigma)\nonumber\\
&&({\tilde g}_{ab}\chi^{\prime b}-{\tilde b}_{ab}\dot{\chi}^b
+\frac{i}{2}\partial_e({\tilde g}_{ab}-{\tilde b}_{ab})
{\tilde \phi}^e_+{\tilde \phi}^b_+-\frac{i}{2}\partial_e
({\tilde g}_{ab}+{\tilde b}_{ab}){\tilde \phi}^e_-
{\tilde \phi}^b_-)|_{\sigma+\pi}=\nonumber\\
&&-({\tilde g}_{ab}\chi^{\prime b}-{\tilde b}_{ab}
\dot{\chi}^b+\frac{i}{2}\partial_e({\tilde g}_{ab}-
{\tilde b}_{ab}){\tilde \phi}^e_+{\tilde \phi}^b_+-
\frac{i}{2}\partial_e({\tilde g}_{ab}+{\tilde b}_{ab})
{\tilde \phi}^e_-{\tilde \phi}^b_-)|_\sigma,
\eea
where the second equation represents that the 
momenta flowing out
of the identification points must have opposite signs,
as in the bosonic
non-abelian case.
The dual fermions satisfy:
\be
\label{z3}
{\tilde \phi}^a_+(\sigma+\pi)=
-i\eta M^{-1}_{ba}M_{bc}{\tilde \phi}^c_-(\sigma).
\ee

\section{Conclusions}
\setcounter{equation}{0}

We have shown that the canonical transformation approach 
is particularly useful to study T-duality in open string 
theories.
It provides the explicit non-local mapping between the target
space coordinates of the original and dual theories. 
{}From it the derivation of the dual
boundary conditions is straightforward.
In backgrounds with abelian isometries we have
reproduced the well-known results of open 
strings - D-branes
duality \cite{ABB,DO}
in a very simple manner.
In the $N=1$ supersymmetric case we have been able to address
arbitrary classical boundary conditions in the initial theory
and study their transformation under abelian T-duality,
showing
that in order to obtain a dual super-D-brane
some conditions over the original backgrounds must be satisfied.
For type I superstrings we have found an orientifold in the 
dual \cite{ABB}.

We have also applied the canonical transformation description of
non-abelian duality to backgrounds of open and closed strings
with non-abelian symmetry groups. In these cases 
the equivalence between initial and dual theories 
at the quantum level is necessary to get generalized
Dirichlet boundary conditions
in the dual, these conditions representing that the 
momenta at the
ends of the string must vanish in a certain curved manifold.
The dual is then a curved $(d-{\rm dim}G-1)$ D-brane, 
static
since the previous conditions of quantum equivalence force
null background gauge field components on the Dirichlet directions.
For unoriented topologies we obtain curved
orientifolds in which the D-branes are hidden.
The $N=1$ supersymmetrization of these theories shows that
the dual is a $N=1$ supersymmetric curved D-brane 
(or orientifold if
we consider unoriented strings).
All these results show that flat D-branes and orientifolds
are just particular features under T-duality.

The canonical transformation description
of T-duality has been very powerful in obtaining information
about the transformation of boundary conditions. 
However we have to remark that some problems are still
open within this approach. 
An explicit proof of the transformation
of the dilaton is lacking, although we believe it should be along
the lines of how the modular anomaly appears in
abelian gauge field theories \cite{Wi,BY}. Also it remains
an open problem the study of
more general examples of non-abelian
backgrounds, namely those with isotropic isometries, like WZW 
models, for
which an explicit canonical transformation description is still not
available. It could be interesting to study the kind of
structures that emerge in the dual.

\subsection*{Acknowledgements}

We would like to thank E. Alvarez, J.L.F. Barb\'on and
A. Gonz\'alez-Ruiz for useful discussions.
Fellowships from Comunidad de Madrid (J.B.) and from M.E.C.
(Spain) (Y.L.) are acknowledged for partial financial support.

\subsection*{Note added}

After this work was completed we learned that in \cite{KS}
curved D-branes were also obtained from Poisson-Lie
T-duality transformations. We would like to thank 
C. Klimcik for bringing this information to our attention
and for very helpful discussions.

\end{document}